\documentclass[showpacs,aps,twocolumn]{revtex4}

\usepackage{bm}
\usepackage{amsmath}
\usepackage{graphicx}
\usepackage{subfigure}
\usepackage[usenames,dvipsnames]{color}
\definecolor{darkblue}{RGB}{0,0,196}
\usepackage[colorlinks=true,linkcolor=darkblue,citecolor=darkblue,urlcolor=darkblue]{hyperref}

\usepackage{setspace}
\usepackage{footmisc}
\usepackage[makeroom]{cancel}
\usepackage{comment}
\usepackage{lineno}

\def\be{\begin{equation}}
\def\ee{\end{equation}}
\def\ba{\begin{eqnarray}}
\def\ea{\end{eqnarray}}

\usepackage{graphicx}
\usepackage{amsmath,bbm}
\usepackage{amssymb,bm}
\usepackage{yfonts}

\begin{document}
\title{Elliptic Flow in Pb+Pb Collisions at $\sqrt{s_{\rm NN}}$ = 2.76 TeV at the LHC Using Boltzmann Transport Equation
with Non-extensive Statistics}
\author{Sushanta Tripathy}
\author{Swatantra Kumar Tiwari}
\author{Mohammed Younus}
\author{Raghunath~Sahoo\footnote{Corresponding author: $Raghunath.Sahoo@cern.ch$}}
\affiliation{Discipline of Physics, School of Basic Sciences, Indian Institute of Technology Indore, Simrol, Indore 453552, India}

\begin{abstract}
\noindent
Elliptic flow in heavy-ion collisions is an important signature of  a possible de-confinement transition from hadronic phase to partonic phase. In the present work, we use non-extensive statistics, which has been used for transverse momentum ($p_{\rm T}$) distribution in proton+proton ($p+p$) collisions, as the initial particle distribution function in Boltzmann Transport Equation (BTE). A Boltzmann-Gibbs Blast Wave (BGBW) function is taken as an equilibrium function to get the final distribution to describe the particle production in heavy-ion collisions. In this formalism, we try to estimate the elliptic flow in Pb+Pb 
collisions at $\sqrt{s_{\rm NN}}$ = 2.76 TeV at the LHC for different centralities.  The elliptic flow ($v_2$) of identified particles 
seems to be described quite well in the available $p_{\rm T}$ range. An approach, which combines the non-extensive nature of particle production in $p+p$ collisions through an evolution in kinetic theory using BTE, with BGBW as an equilibrium distribution is successful in describing
the spectra and elliptic flow in heavy-ion collisions.

\pacs{25.75-q,12.38.Mh, 25.75.Ld, 25.75.Dw}

\end{abstract}
\date{\today}
\maketitle 
\section{Introduction}
\label{intro}
The early Universe has always been a mystery and understanding it gives a major challenge to the scientists from all fields. The relativistic heavy-ion collision experiments at STAR@RHIC-BNL, ALICE@LHC-CERN etc., provide us with an unique but brief opportunity to peer into a system that looks very much like early universe, a cauldron of de-confined quarks and gluons at an extreme temperature and/or energy density. Heavy-ion collisions at relativistic energies form a state which is a short lived, thermalized system of quarks and gluons, and are de-confined within a volume larger than nucleonic size but smaller than a nucleus. This thermalized system is called quark-gluon-plasma (QGP) and it provides us with some unique signatures like jet quenching, elliptic flow etc.~\cite{Younus:2010sx,Younus:2011mn,Esumi:2002vy,Lin:1994xma}.

Elliptic flow is particularly intriguing as it is believed to be generated at the earliest phase of QGP when partons within the system undergo multiple interaction and try to attain transverse momentum ($p_{\rm T}$) isotropization in azimuthal plane~\cite{Sorge:1998mk}. In other words, azimuthal anisotropy or elliptic flow of the system, measures the remaining asymmetry of particle density in momentum space relative to the reaction plane after hadronization. In non-central heavy-ion collisions, where the impact parameter is non-zero for each event, the overlap region resembles an almond or oval shape with the major axis perpendicular to the reaction plane and its length in and out of the plane would be different~\cite{Sun:2014rda}. Thus the particles moving along different azimuthal direction would exhibit angular dependent property in their spectra. When the system evolves, the anisotropy in the co-ordinate space is manifested as the momentum space anisotropy due to the difference in pressure gradients along different azimuthal directions. Speaking in detail, both gluons and light quarks exhibit this flow pattern because of the multiple scattering they undergo, and consequently early thermalization for the entire system of partons sets in, as predicted earlier by theoretical models
~\cite{Bjorken:1982qr,Kapusta:1992uy,Huovinen:2001cy}. It can be shown that the bulk of the partons at low and medium momentum regions exhibit this collective behaviour which is experimentally observed as one of the prominent signatures of partons' de-confinement and the formation of QGP. If the thermalized system expands rapidly and cools, a part of the information on initial geometrical anisotropy will be deluged. However, experimental results have precise data of flow being measured for identified hadrons, heavy mesons etc. for the entire range of $p_{\rm T}$~\cite{Adams:2005dq, Adcox:2004mh}. Thus, it is suggested that the dynamics at the freeze-out surfaces may affect the particle flow. A competition between relaxation time, kinetic and chemical freeze-out should play a vital role in the observed particle spectra and their elliptic flow~\cite{Retiere:2003kf,Hirano:2005xf}. Phenomenological study of these important effects on measured hadrons' flow have developed the idea into an interesting contemporary topic~\cite{Das:2017dsh}. In the present article, we have tried to develop the calculations of $v_2$ of identified hadrons using transport formalisms. We will return to this topic in detail later in our paper. 

Now, the particle distribution in four-momentum space can be written as a Fourier series~\cite{Sun:2014rda},

\ba
 \label{eq1}
  E\frac{d^3N}{dp^3}=\frac{1}{2\pi}\frac{d^{2}N}{p_{T}dp_{T}dy}\left(1+2\sum_{n=1}^{\infty} v_{n}\,\cos(n\phi)\right),
\ea

where $E$ is the energy and $y$ is the rapidity of produced particles, $v_n$ is the $n-th$ harmonic co-efficient of the flow and $\phi$ is the azimuthal angle of a particle. The second harmonic co-efficient is called elliptic flow ($v_2$) of the system, which signifies the transformation from geometric asymmetry to momentum space asymmetry due to the strong interaction of quarks and gluons.

Earlier, many theoretical calculations which are based on transport equations \cite{Kolb:2000fha,Das:2017dsh} and phenomenological models \cite{Younus:2012yi} have successfully explained $v_2$. In the present article, for the first time we explain $v_2$ using non-extensive statistics in BTE making a direct connection of particle production in
$p+p$ collisions to elliptic flow in heavy-ion collisions.

In the following sections we will present two major components adapted in our calculations: 1) Non-extensive Tsallis statistics and its effect on final elliptic flow, 2) Use of Boltzmann transport equation (BTE) in the evolution of particle momentum distribution in a thermal medium. Later, we will present our results along with the discussion of our findings, and ultimately a summary of this article will follow.

\noindent
\section{Non-extensive Tsallis statistics}
\label{tsallis1}
Let us now move onto a brief discussion of the non-extensivity in our calculations. Observation of huge multiplicities of hadrons in experiments at RHIC and LHC suggests formation of a thermalized system in the earlier stages of heavy-ion collisions. Thus statistical approaches to deal with such systems are more suitable for explaining the data. However, it is possible that the produced matter may be slightly deviant from such locally equilibrated state. As such, Tsallis statistics may give a better explanation of such non-equilibrium systems~\cite{Cleymans:2011in,Thakur:2016boy}. It is known that Boltzmann-Gibbs (BG) statistics which is meant for system at thermal equilibrium, can explain a fully equilibrated system incorporating various modifications at freeze-out surfaces and their parameters. On the other hand, Tsallis statistics with its non-extensivity can be regarded as a generalization of exponential BG statistics and gives a better explanation of systems which have not yet reached equilibration. It features a power-law like structure and a non-extensive, entropic $q$-parameter which shows the extent of non-equilibration of any particle in a thermal bath. The $p_{\rm T}$-spectra of the identified particles in $p+p$ collisions at the LHC are very well described by Tsallis non-extensive statistics \cite{Cleymans:2011in}. In addition, it has been observed that an increase in particle multiplicity drives a system towards thermodynamic equilibrium~\cite{Khuntia:2017ite}.

The transverse momentum spectra can be described by a thermodynamically consistent Tsallis non-extensive distribution function as~\cite{Cleymans:2011in},
\ba
\label{eq2}
f_{q}(p_T)=C_q\left[1+(q-1)\,\frac{p_T}{T_{ts}} \right]^{\frac{-q}{q-1}}\,,
\ea
where $C_q$ is the normalization, $T_{ts}$ is the Tsallis temperature and $q$ is the Tsallis non-extensive parameter. 

The Tsallis distribution converges to the usual Boltzmann-Gibbs (BG) distribution when the $q$-parameter goes to unity:

\ba
\label{eq3}
f(p_T)|_{q\rightarrow 1}=C_1 exp\left(-\frac{p_T}{T}\right)\,,
\ea
where $T$ is the equilibrium temperature.

As shown in earlier publications~\cite{Tripathy:2016hlg,Tripathy:2017kwb} that Tsallis statistics have been able to explain particle spectra and nuclear modification factor ($R_{AA}$), when used as input to the Boltzmann transport equation(BTE). Next we move onto our following section where we discuss BTE with Tsallis distribution as input and calculate identified hadrons' elliptic flow ($v_2$)~\cite{Abelev:2014pua}.

\noindent
\section{Elliptic Flow in Relaxation time approximation (RTA) of Boltzmann transport equation (BTE)}
\label{raa}
Let us briefly revisit certain points we mentioned in the introductory section. The effects of the particle evolution within the medium as well as effects due to dynamics at the freeze-out surfaces have profound effects on the final particle spectra. The transport calculations such as hydrodynamics, BTE etc. are most suitable for determining these effects both qualitatively as well as in a quantitative manner~\cite{Baier:2006um,Gavin:1985ph,Geiger:1991nj,Srivastava:1997qg,Bass:2004vh,Zhang:1999rs,Younus:2013rja}. We know that various dynamical features ranging from multi-parton interaction, in-medium energy loss, thermal, and chemical equilibrations, and dynamics at freeze-out surfaces contribute extensively to the particle flow and can be studied using BTE. The transport properties and their numerical estimation bring forth, many in-depth information about the observed hadron spectra, ratios, azimuthal anisotropy etc.

The BTE in general can be written as:
\begin{eqnarray}
\label{eq2}
 \frac{df(x,p,t)}{dt}=\frac{\partial f}{\partial t}+\vec{v}.\nabla_x
f+\vec{F}.\nabla_p
f=C[f],
\end{eqnarray}
where $f(x,p,t)$ is the distribution of particles which depends on position, momentum and time. $\vec{v}$ is the velocity and $\vec{F}$ is the external force. $\nabla_x$ and $\nabla_p$ are the partial derivatives with respect to position and momentum, respectively. $C[f]$ is the collision term which encodes the interaction of the probe particles with the medium. Earlier, BTE has also been used in RTA to study the time evolution of temperature fluctuation in a non-equilibrated system \cite{Bhattacharyya:2015nwa} and also for studying the $R_{AA}$ of various light and heavy flavours at RHIC and LHC energies \cite{Tripathy:2016hlg}. 

Assuming homogeneity of the system ($\nabla_x f=0$) and in absence of external forces ($\vec{F}=$0), the second and third terms in Eq. \ref{eq2} become zero and the equation reduces to,
\ba
 \label{eq3}
  \frac{df(x,p,t)}{dt}=\frac{\partial f}{\partial t}=C[f].
\ea

In RTA \cite{Florkowski:2016qig}, the collision term is expressed as:
\ba
\label{eq4}
 C[f] =-\frac{f-f_{eq}}{\tau},
 \label{colltermrta}
\ea
where $f_{eq}$ is Boltzmann local equilibrium distribution characterized by a temperature $T$. $\tau$ is the relaxation time, the time taken by a non-equilibrium system to reach equilibrium. Using Eq. \ref{colltermrta}, Eq. \ref{eq3} becomes 
\ba
 \label{eq5}
  \frac{\partial f}{\partial t}=-\frac{f-f_{eq}}{\tau}.
\ea
Solving the above equation with the initial conditions {\it i.e.} at $t=0, f=f_{in}$ and at $t=t_f, f=f_{fin}$, we get,
\ba
 \label{eq6}
 f_{fin}=f_{eq}+(f_{in}-f_{eq})e^{-\frac{t_f}{\tau}},
\ea
where $t_f$ is the freeze-out time. We use Eq. \ref{eq6} in the definition of the elliptic flow ($v_2$) at mid-rapidity, which is expressed as,
\ba
\label{eq7}
v_2(p_T)=\frac{\int{f_{fin} \times \cos(2\phi)\,d\phi}}{\int{f_{fin}\,d\phi}}.
\ea
Eq. \ref{eq7} gives azimuthal anisotropy after incorporating RTA in BTE. It involves the Tsallis non-extensive distribution function as the initial distribution of particles and BGBW function as the equilibrium distribution. Here, we take Boltzmann-Gibbs Blast Wave (BGBW) function, $f_{eq}$, as \cite{Cooper}: 
\ba
\label{bgbw1}
f_{eq}=D \int d^3\sigma_\mu p^\mu exp(-\frac{p^\mu u_\mu}{T})\,,
\ea
where the particle four-momentum is, $p^\mu = (m_T\cosh y, p_T\cos\phi, p_T\sin\phi, m_T\sinh y)$, the four-velocity denoting flow velocities in space-time is given by, $u^\mu = \cosh\rho(\cosh\eta, \tanh\rho \cos\phi_r, \tanh\rho \sin \phi_r, \sinh \eta)$, while the kinetic freeze-out surface is given by $d^3\sigma_\mu = (\cosh\eta, 0, 0, -\sinh\eta)\tau rdrd\eta d\phi_r$. Here, $\eta$ is the space-time rapidity. With simplification, assuming Bjorken correlation in rapidity, $i.e.$ $y=\eta$~\cite{Bjorken:1982qr}, Eq.~\ref{bgbw1} can be expressed as:  
\ba
\label{boltz_blast}
f_{eq} = D \int_0^{R_{0}} r\;dr\;K_1\Big(\frac{m_T\;\cosh\rho}{T}\Big)I_0\Big(\frac{p_T\;\sinh\rho}{T}\Big),
\ea
where $D = \displaystyle \frac{gVm_T}{2\pi^2}$. Here $g$ is the degeneracy factor, $V$ is the system volume, and $m_{\rm T}=\sqrt{p_T^2+m^2}$ is the transverse mass.
Here $K_{1}\displaystyle\Big(\frac{m_T\;{\cosh}\rho}{T}\Big)$ and $I_0\displaystyle\Big(\frac{p_T\;{\sinh}\rho}{T}\Big)$ are the modified Bessel's functions and are given by
\begin{widetext}
\ba
\centering
K_1\Big(\frac{m_T\;{\cosh}\rho}{T}\Big)=\int_0^{\infty} {\cosh}y\;{\exp}\Big(-\frac{m_T\;{\cosh}y\;{\cosh}\rho}{T}\Big)dy,
\ea
\ba
\centering
I_0\Big(\frac{p_T\;{\sinh}\rho}{T}\Big)=\frac{1}{2\pi}\int_0^{2\pi} exp\Big(\frac{p_T\;{\sinh}\rho\;{\cos}\phi}{T}\Big)d\phi,
\ea
\end{widetext}
where $\rho$ in the integrand is a parameter given by $\rho=tanh^{-1}\beta_r+\rho_a \cos(2\phi)$, with $\rho_a$ as the anisotropy parameter in the flow, and $\beta_r=\displaystyle\beta_s\;\Big(\xi\Big)^n$ \cite{Huovinen:2001cy,Schnedermann:1993ws,BraunMunzinger:1994xr, Tang:2011xq} is the radial flow. $\beta_s$ is the maximum surface velocity and $\xi=\displaystyle\Big(r/R_0\Big)$, with $r$ as the radial distance. In the blast-wave model the particles closer to the center of the fireball move slower than the ones at the edges. The average of the transverse velocity can be evaluated as \cite{Adcox:2003nr}, 
\ba
<\beta_r> =\frac{\int \beta_s\xi^n\xi\;d\xi}{\int \xi\;d\xi}=\Big(\frac{2}{2+n}\Big)\beta_s.
\ea
In our calculation we use a linear velocity profile, ($n=1$) and $R_0$ is the maximum radius of the expanding source at freeze-out ($0<\xi<1$). In this analysis, the initial distribution is the parameterized Tsallis distribution~\cite{Cleymans:2011in}
\ba
\label{eq13}
f_{in}=\frac{D}{2}
\left[1+{(q-1)}{\frac{m_T}{T_{ts}}}\right]^{-\frac{q}{q-1}}.
\ea
Thus, we have used the Tsallis distribution to obtain the expression for the final distribution and elliptic flow, $v_2$. The thermodynamically consistent Tsallis distribution is used for studying the particle distributions stemming from the proton-proton collisions as discussed in Ref. \cite{Cleymans:2011in}.
Using Eqs. \ref{boltz_blast} and \ref{eq13}, the final distribution can be expressed as, 


\begin{widetext}
\ba
\label{eq14}
\nonumber f_{fin} = D \Bigg[\int_0^{R_{0}} r\;dr\;K_1\Big(\frac{m_T\;{\cosh}\rho}{T}\Big)I_0\Big(\frac{p_T\;{\sinh}\rho}{T}\Big)+\\\left(\frac{1}{2}\left[1+{(q-1)}{\frac{m_T}{T_{ts}}}\right]^{-\frac{q}{q-1}}-\int_0^{R_{0}}r\;dr\;K_1\Big(\frac{m_T\;{\cosh}\rho}{T}\Big)I_0\Big(\frac{p_T\;{\sinh}\rho}{T}\Big)\right)e^\frac{-t_{f}}{\tau}\Bigg].
\ea
\end{widetext}
Using Eq. \ref{eq14} (both for mid-rapidity and for zero chemical potential) in Eq. \ref{eq7}, we calculate $v_2$ for the observed identified hadrons as follows:

\ba
\label{eq15}
 v_{2}(p_T) =\frac{P}{Q},
\ea
where
\begin{widetext}
\ba
P&=&D\,\int d\phi \cos(2\phi)\Bigg[\int_0^{R_{0}} r\;dr\;K_1\Big(\frac{m_T\;{\cosh}\rho}{T}\Big)I_0\Big(\frac{p_T\;{\sinh}\rho}{T}\Big)\nonumber\\
&+&\left(\frac{1}{2}\left[1+{(q-1)}{\frac{m_T}{T_{ts}}}\right]^{-\frac{q}{q-1}}-\int_0^{R_{0}}r\;dr\;K_1\Big(\frac{m_T\;{\cosh}\rho}{T}\Big)I_0\Big(\frac{p_T\;{\sinh}\rho}{T}\Big)\right)e^\frac{-t_{f}}{\tau}\Bigg]\,,
\ea
\end{widetext}

\begin{widetext}
\ba
Q&=&D\,\int d\phi \Bigg[\int_0^{R_{0}} r\;dr\;K_1\Big(\frac{m_T\;{\cosh}\rho}{T}\Big)I_0\Big(\frac{p_T\;{\sinh}\rho}{T}\Big)\nonumber\\
&+&\left(\frac{1}{2}\left[1+{(q-1)}{\frac{m_T}{T_{ts}}}\right]^{-\frac{q}{q-1}}-\int_0^{R_{0}}r\;dr\;K_1\Big(\frac{m_T\;{\cosh}\rho}{T}\Big)I_0\Big(\frac{p_T\;{\sinh}\rho}{T}\Big)\right)e^\frac{-t_{f}}{\tau}\Bigg]\,.
\ea
\end{widetext}


The present formalism of elliptic flow incorporates the non-extensive nature of particle productions in $p+p$ collisions through a kinetic theory in BTE with RTA and relates to azimuthal anisotropy in LHC. 

Next we move to results and discussion to see how effectively the present formalism is successful in describing the elliptic flow at the LHC energies.
 
\begin{figure}[t]
\includegraphics[height=21em]{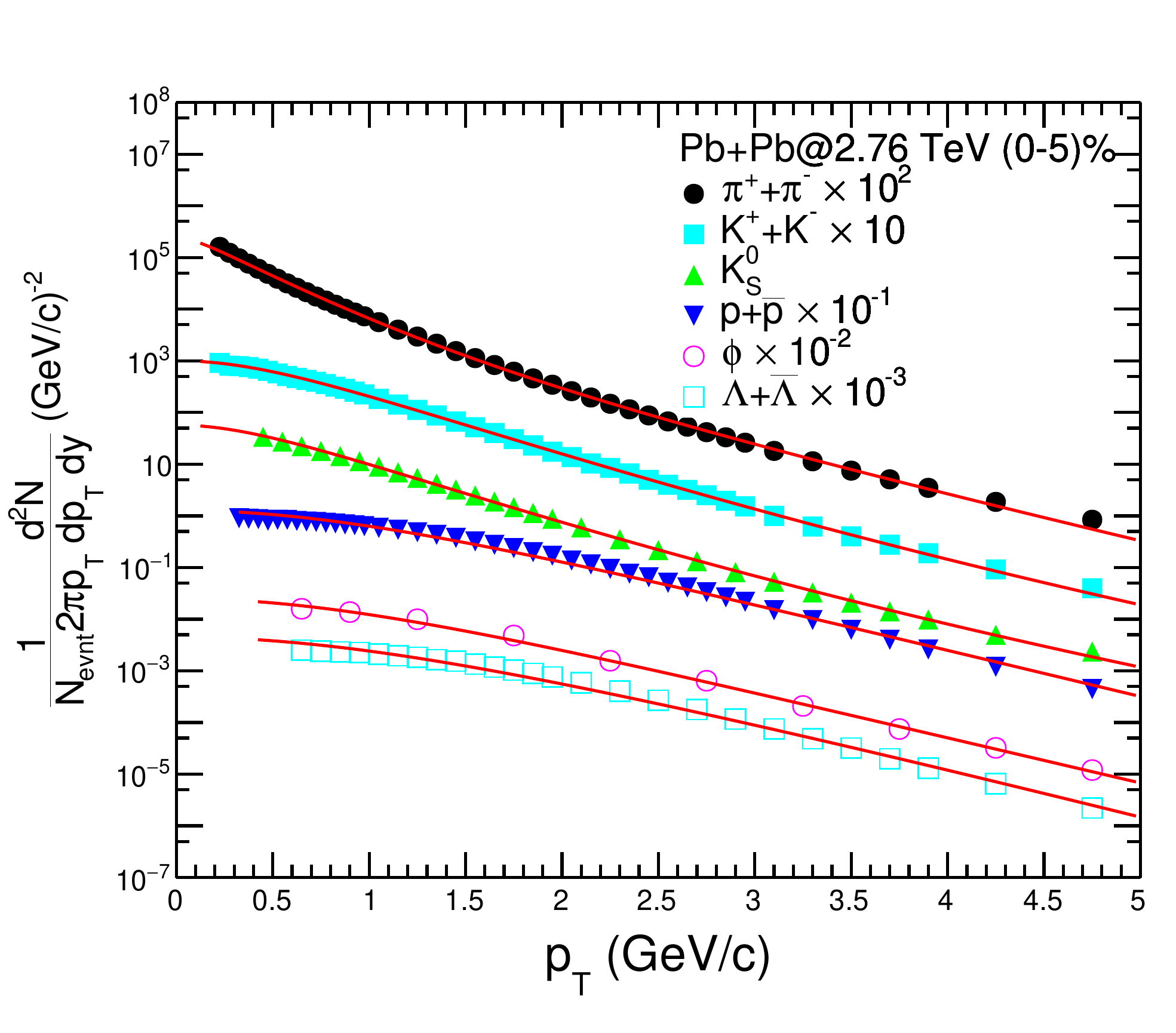}
\caption[]{(Color online) Transverse momentum distribution of identified hadrons for 0--5\% centrality in Pb+Pb collisions at $\sqrt{s_{\rm NN}}$ = 2.76 TeV~\cite{Abelev:2014laa,Adam:2017zbf,Abelev:2013xaa}. Eq. \ref{eq14} is fitted to the spectra of the identified particles.}
\label{fig1}
\end{figure}
    
\section{Results and Discussions}
\label{results}

\begin{figure}[h]
\includegraphics[height=21em]{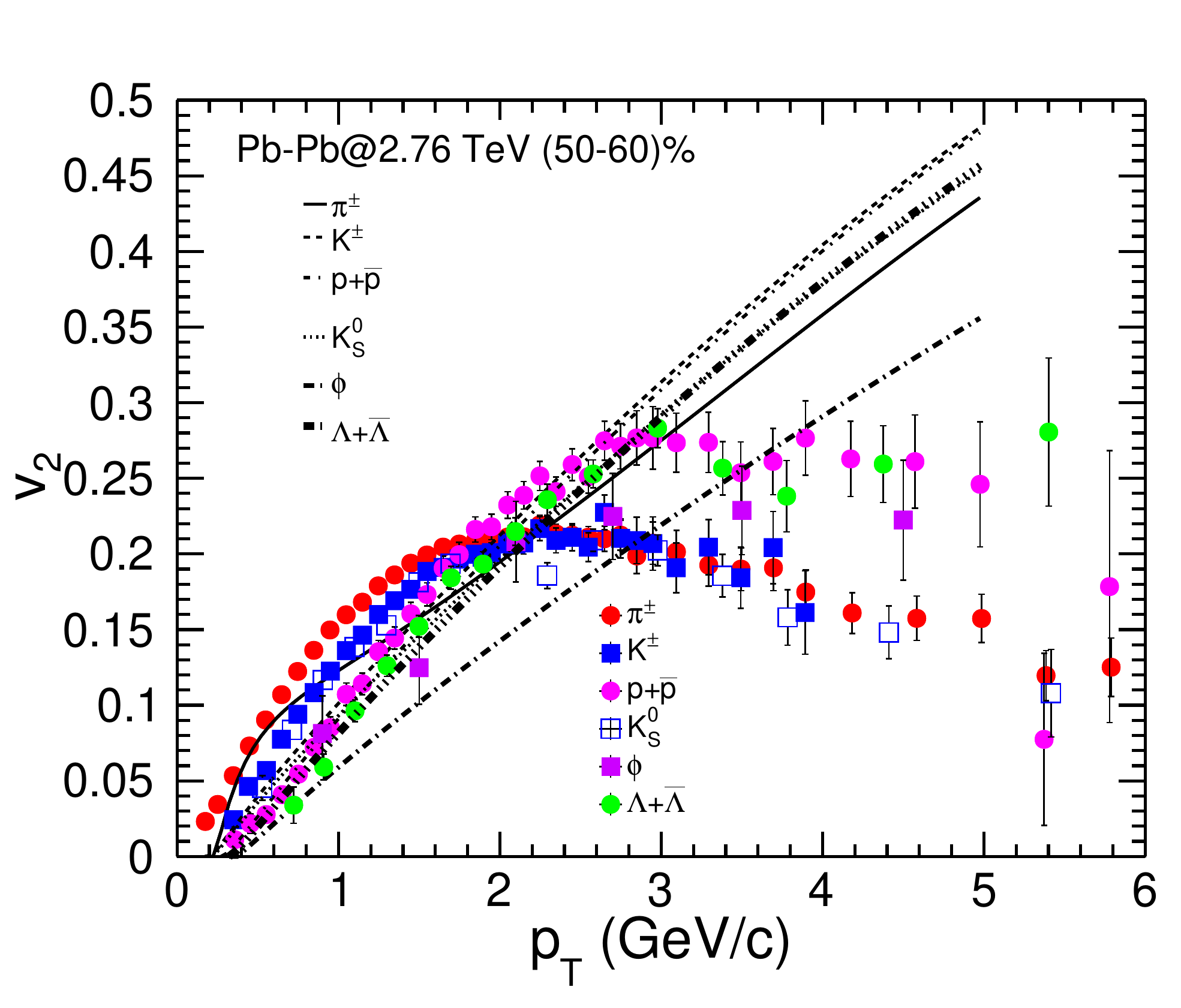}
\caption[]{(Color online) Elliptic flow of identified hadrons for 50-60\% centrality in Pb+Pb collisions at $\sqrt{s_{\rm NN}}$ = 2.76 TeV using only BGBW 
(Eq. \ref{boltz_blast}).}
\label{fig2}
\end{figure}

We now proceed to the more detailed analysis of the experimental data. We have fixed the kinetic freeze-out temperature for central collisions at 0.095 GeV and that of peripheral collisions at 0.11 GeV~\cite{Abelev:2013vea}. Keeping the rest of the parameters free, we fit the experimental data using the TMinuit class available in the ROOT library \cite{root} to get a convergent solution. The convergent solution is obtained by $\chi ^2$-minimization technique. 

\begin{figure}
\includegraphics[height=21em]{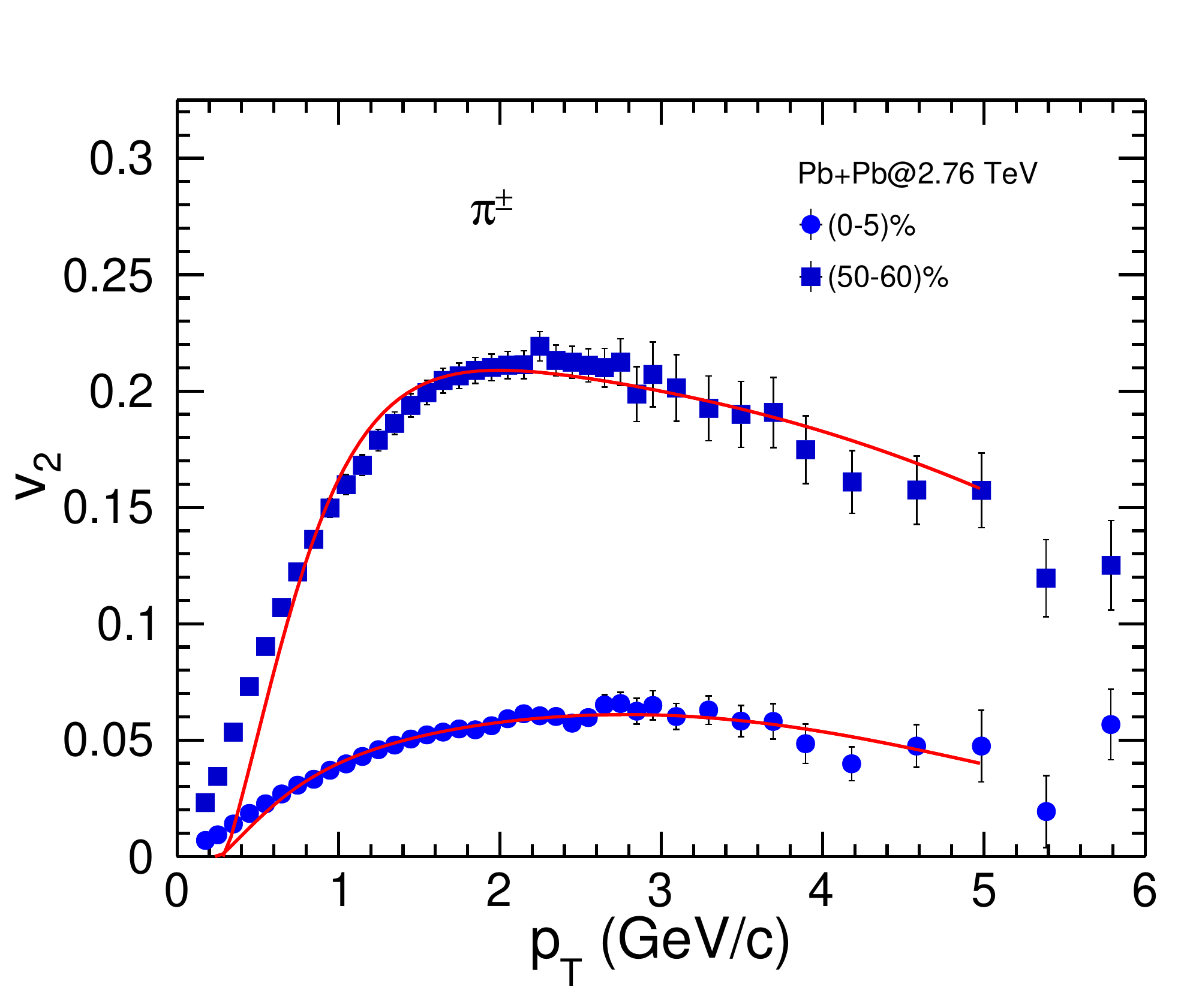}
\caption[]{(Color online) Elliptic flow of $\pi^{\pm}$ meson for (0-5)\% and (50-60)\% centralities~\cite{Abelev:2014pua}.}
\label{fig3}
\end{figure}

\begin{figure}
\includegraphics[height=21em]{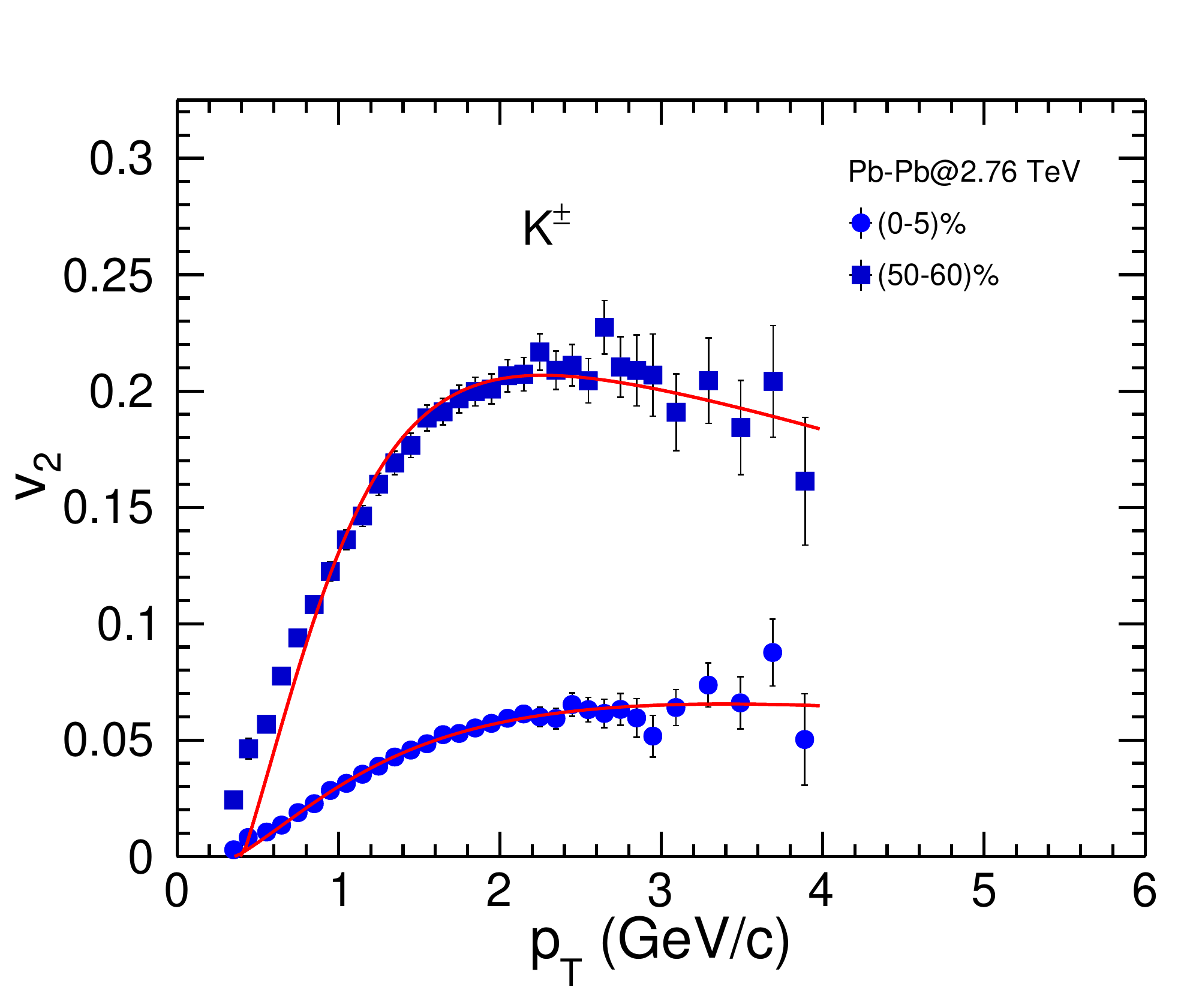}
\caption[]{(Color online)Elliptic flow of $K^{\pm}$, kaon for (0-5)\% and (50-60)\% centralities~\cite{Abelev:2014pua}. }
\label{fig4}
\end{figure}

\begin{figure}
\includegraphics[height=21em]{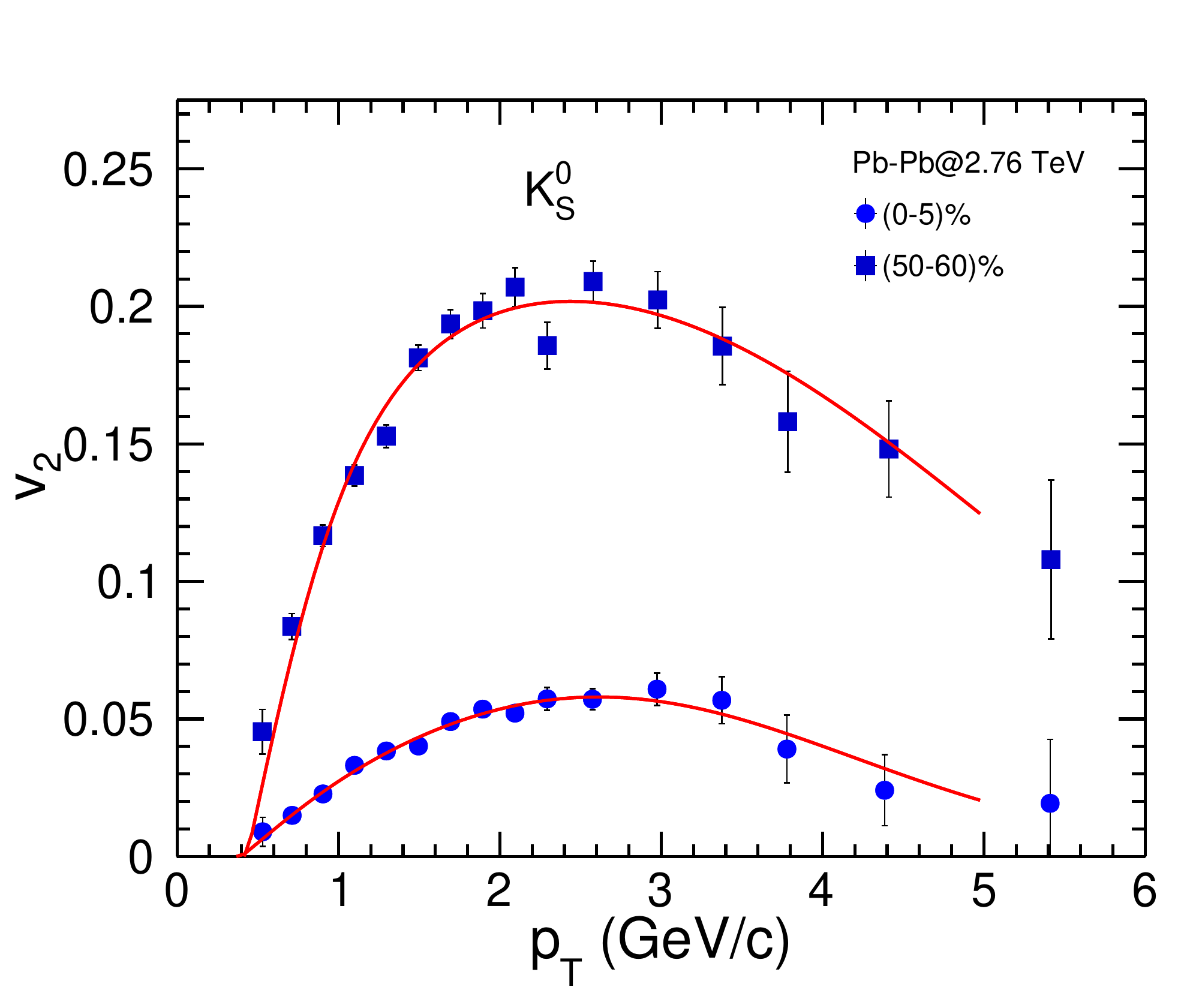}
\caption[]{(Color online) Elliptic flow of $K_{S}^{0}$ for (0-5)\% and (50-60)\% centralities~\cite{Abelev:2014pua}.}
\label{fig5}
\end{figure}

\begin{figure}
\includegraphics[height=21em]{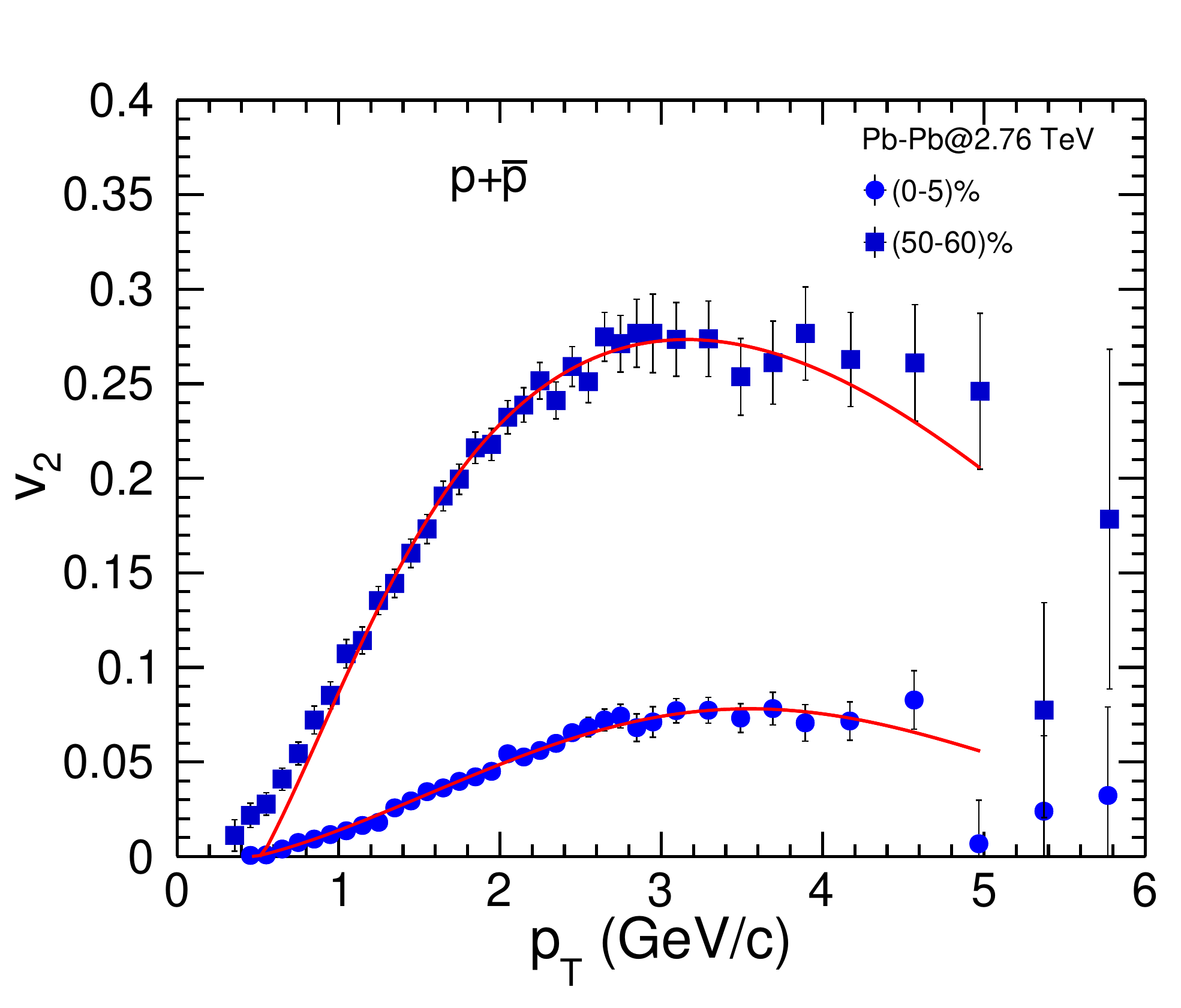}
\caption[]{(Color online) Elliptic flow of protons for (0-5)\% and (50-60)\% centralities~\cite{Abelev:2014pua}.}
\label{fig6}
\end{figure}

\begin{figure}
\includegraphics[height=21em]{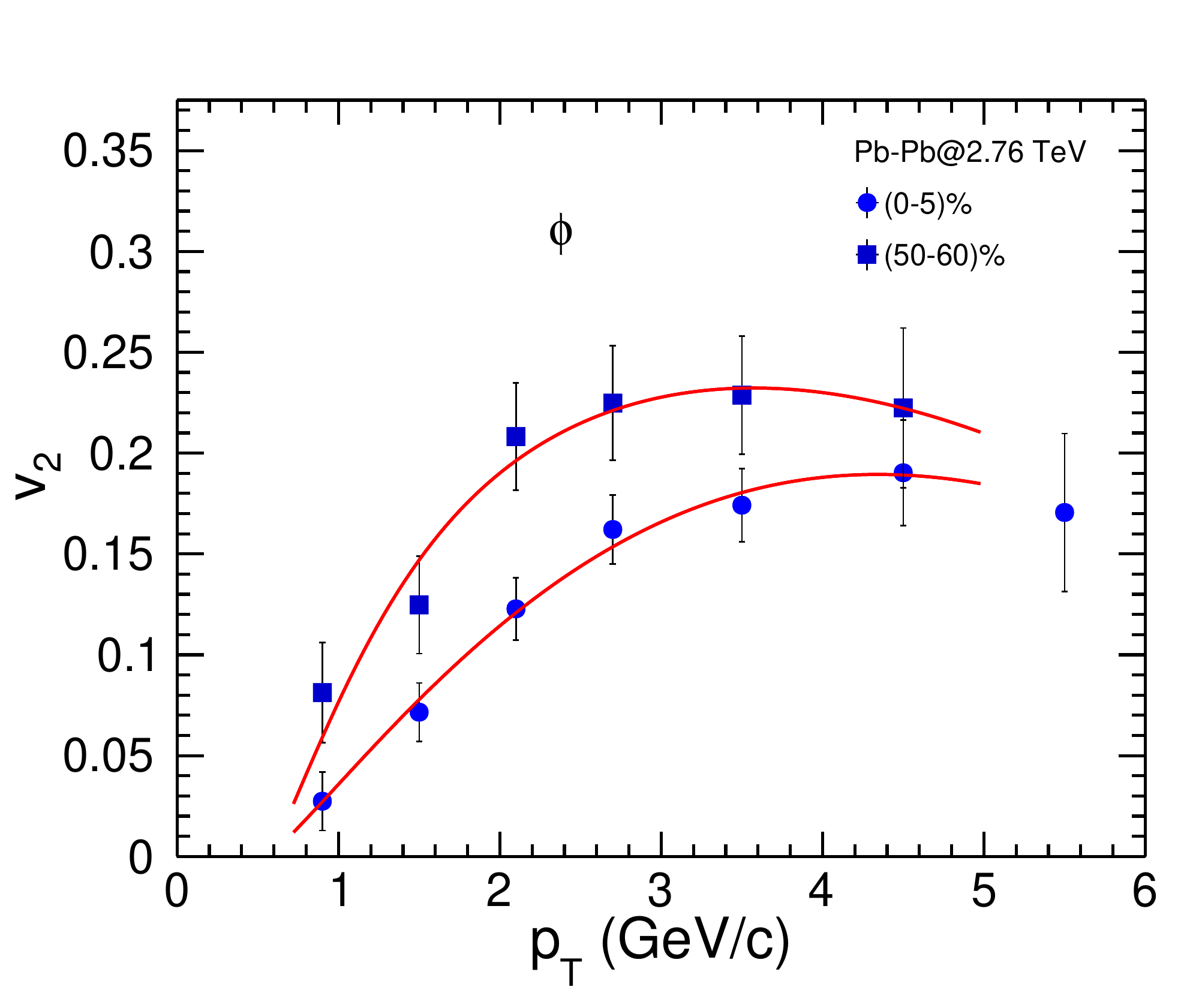}
\caption[]{(Color online) Elliptic flow of $\phi$ meson for (0-5)\% and (50-60)\% centralities~\cite{Abelev:2014pua}.}
\label{fig7}
\end{figure}

\begin{figure}
\includegraphics[height=21em]{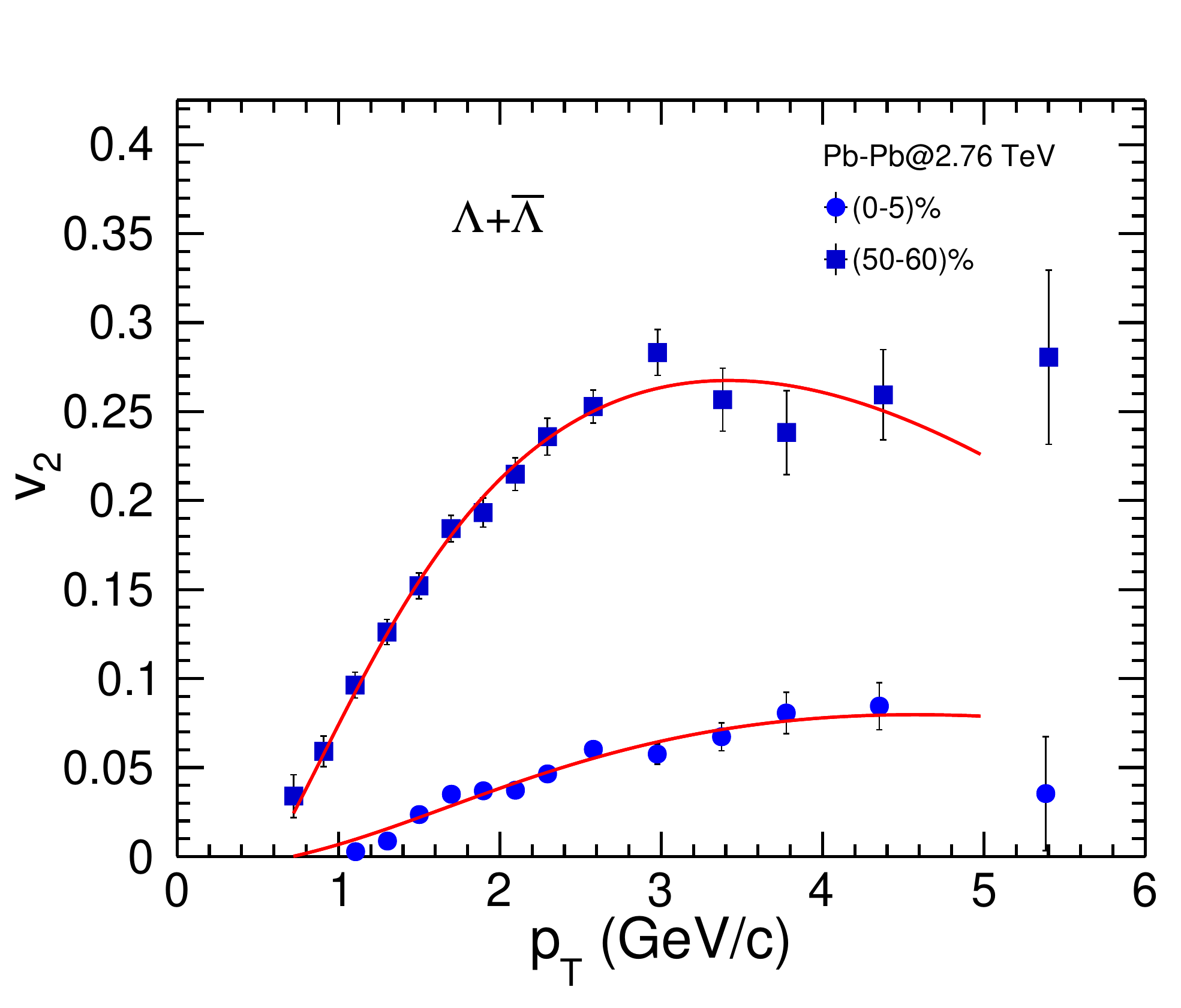}
\caption[]{(Color online) Elliptic flow of $\Lambda$ hyperon for (0-5)\% and (50-60)\% centralities~\cite{Abelev:2014pua}.}
\label{fig8}
\end{figure}

Fig.~\ref{fig1} shows $p_{\rm T}$-spectra for available identified particles for 0-5\% centrality in Pb+Pb collisions at $\sqrt{s_{\rm NN}}$ = 2.76 TeV~\cite{Abelev:2014laa,Adam:2017zbf,Abelev:2013xaa}. Eq. \ref{eq14} is fitted to the $p_{\rm T}$-spectra of different particles, which shows a very good description using the present formalism.
From Figs.~\ref{fig2} to \ref{fig8}, we have shown elliptic flow ($v_2$) for identified hadrons ($\pi,K,p,K_S^0,\phi,\Lambda$)~\cite{Abelev:2014pua} using Eq.~\ref{eq15}. The plots are shown for two different centralities, (0-5)\% and (50-60)\%.

Fig.~\ref{fig2} shows elliptic flow of hadrons using only BGBW formalism given by Eq. \ref{boltz_blast}. BGBW formalism is valid when the system goes to complete local thermal equilibrium. In this figure, although we find that BGBW agrees with experimental data at low-$p_T$, it deviates completely from data points beyond $p_T>$ 2 GeV/c. 
This indicates that particles with lower momenta seem to show a tendency of equilibration. To have a complete description of $v_2$ in the available $p_{\rm T}$-range, we proceed to use the present formalism in its full form.

From Fig.~\ref{fig3} to Fig.~\ref{fig8}, we have $v_2$ for hadrons using Tsallis distribution as input and BGBW as equilibrium distribution in BTE. The parametrization for the BGBW is taken at kinetic freeze-out surface which is identical to that of freeze-out in (2+1)-d hydrodynamical calculations. The $v_2$ plots are for most central (0-5\%) and peripheral (50-60\%) Pb+Pb collisions at $\sqrt{s_{\rm NN}}$ = 2.76 TeV. For all of these, we have used Eq.~\ref{eq15} for the description of elliptic flow using the present formalism. We observe that the BTE with Tsallis distribution could explain the elliptic flow for all the measured hadrons at both the centralities. The $v_2$ curve rises till $p_T\leq$ 2 GeV/c and then it tends to fall, which shows that particles in these two different regions of $p_{\rm T}$, have a distinct and opposite medium effect on their flow pattern. Comparing with a simple BGBW formalism, as shown in Fig. \ref{fig2}, one may infer that the particles with lower momenta have a tendency of equilibration. However as mentioned earlier, the freeze-out dynamics play a vital role in determining the flow pattern of any particle. This is evident from Figs.~\ref{fig3}-\ref{fig8}. The flow pattern included at the freeze-out surface gives the characteristic increase in the elliptic flow at low $p_{\rm T}$ region whereas, the non-extensivity in the initial distribution has profound effect on the particles at the mid and high $p_{\rm T}$. The detailed study of the set of parameters should give us an indirect view of physics behind such a flow pattern and the extent of equilibration, effects of freeze-out dynamics on final hadron spectra. Let us now move to our next set of results where we would discuss the parameters in details.

\begin{figure}
\includegraphics[height=21em]{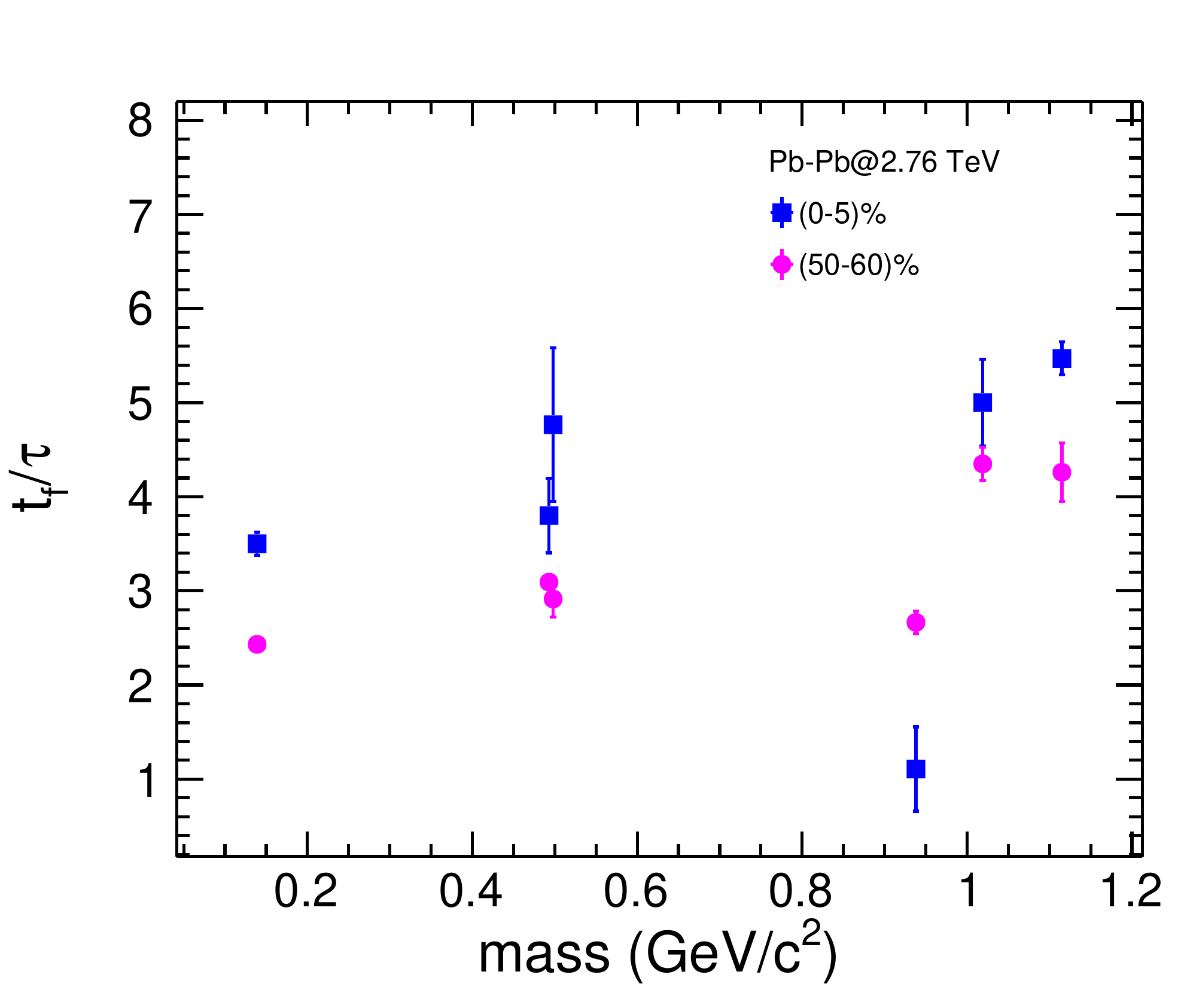}
\caption[]{(Color online) The ratio of freeze-out time to relaxation time, $t_f/\tau$ for different identified hadrons.}
\label{fig9}
\end{figure}

In Fig.~\ref{fig9}, we have shown the parameter, $t_f/\tau$ vs mass, $m$ of the identified hadrons. The two sets in the figure are for two different centralities. While $\tau$ depicts relaxation time or equilibration time for the particle, $t_f$ is the time when the particle finally stops interacting. This particular ratio of parameters is interesting as it shows an interplay between relaxation and freeze-out time. The assumption in our calculation is that the equilibration time for the light flavours should always be less than kinetic freeze-out time, so that the ratio, $t_f/\tau$ is always greater than unity. Similarly, for the heavy quarks or jet particles with lower degree of drag and diffusion, this ratio might be less than unity~\cite{Moore:2004tg,Akamatsu:2008ge}. 
The plot shows that the parameter, $t_f/\tau$ increases with the mass of the particle. However, the rise in the values of $t_f/\tau$ with mass or the slope is more in central collisions as compared to peripheral collisions. This suggests more rapid equilibration for the particles at the central collision. In the present calculations, the hadrons having larger mass undergo a decrease in $\tau$ or an increase in $t_f$. In either case the ratio should increase with the mass as seen in our results. 
This leaves out further investigations to have a clear understanding of the interplay between thermalization and freeze-out. 

\begin{figure}
\includegraphics[height=21em]{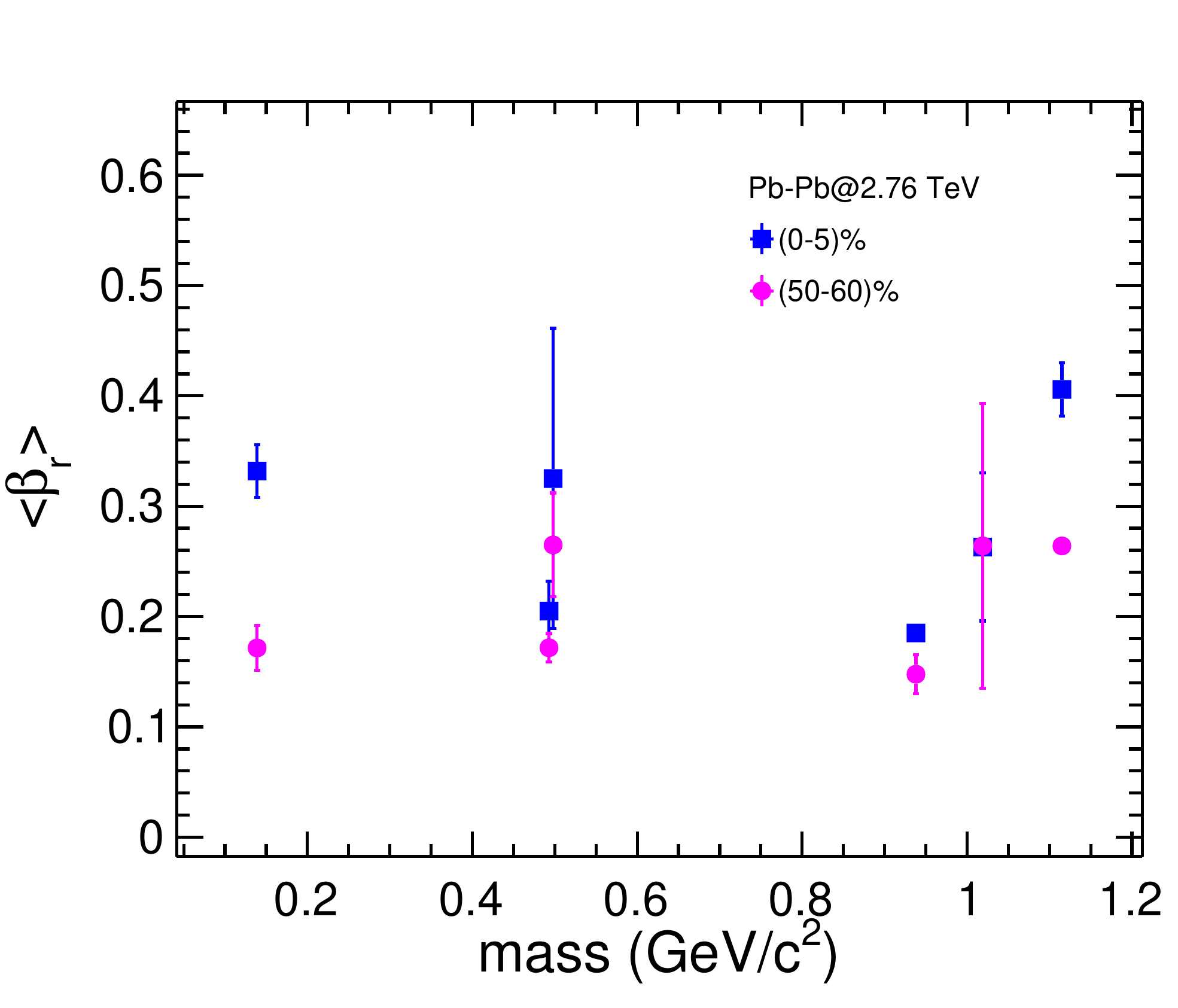}
\caption[]{(Color online) Radial flow parameter, $\langle\beta_r\rangle$ for identified hadrons.}
\label{fig10}
\end{figure}

In Fig.~\ref{fig10}, we have shown the dependency of radial flow parameter, $\langle\beta_r\rangle$ on the mass of hadrons. In our radial flow profile, we have used $n$=1 assuming linear increase in the radial flow velocity. The radial flow parameter does not show a clear mass dependence, which hints for a collectivity in the system. However, the values for central collisions are higher than that of peripheral collisions. This goes inline with the earlier observations on centrality dependence of radial flow \cite{Abelev:2013vea}. 

\begin{figure}
\includegraphics[height=21em]{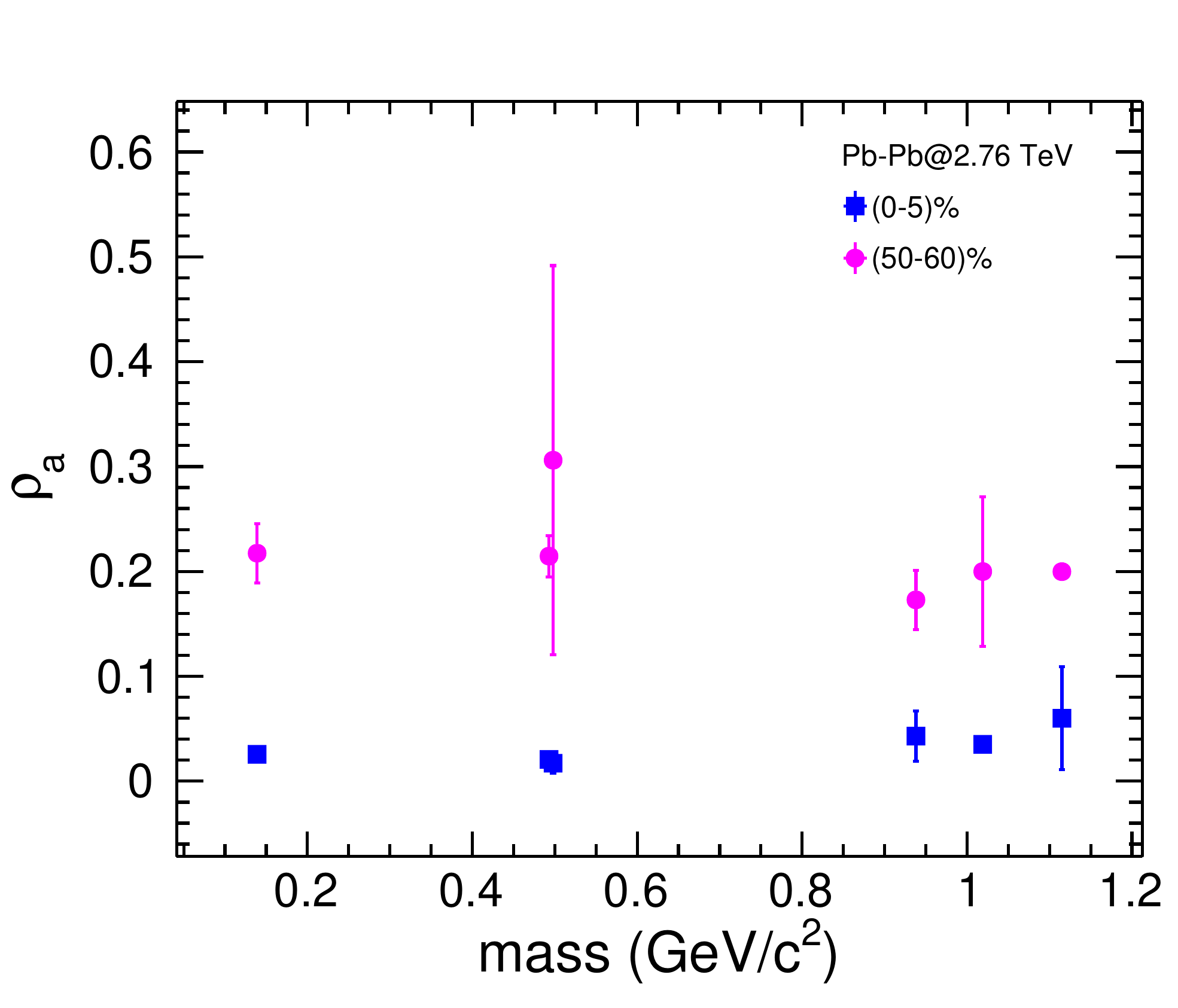}
\caption[]{(Color online) Anisotropy parameter, $\rho_a$ for identified hadrons.}
\label{fig11}
\end{figure}

In Fig.~\ref{fig11}, we have shown the dependence of anisotropy parameter, $\rho_a$ on mass of the hadrons. Apart from $\langle\beta_r\rangle$ discussed in last paragraph, $\rho_a$ is also important as it shows the extent of anisotropy embedded in the transverse rapidity, $\rho$. For the central collisions (0-5\%), the parameter values do not change much with the mass, while for peripheral collisions, $\rho_a$ seems mass independent. Although this is clearly visible in the experimental data (see Fig. \ref{fig2}) up to $p_T \sim$ 2 GeV/c, the $p_{\rm T}$-dependent dynamics dominated by effects like recombination, jet-events, etc. take over for higher momentum range.  In addition, it is evident that with centrality the anisotropy in the flow increases, which is expected. To make this statement more evident, the flow in general, could be understood as an interplay
of radial flow and the elliptic flow, which arises from initial state momentum anisotropy. When the radial flow is higher in central collisions compared to the peripheral ones, the elliptic flow behaves differently.

\begin{figure}
\includegraphics[height=21em]{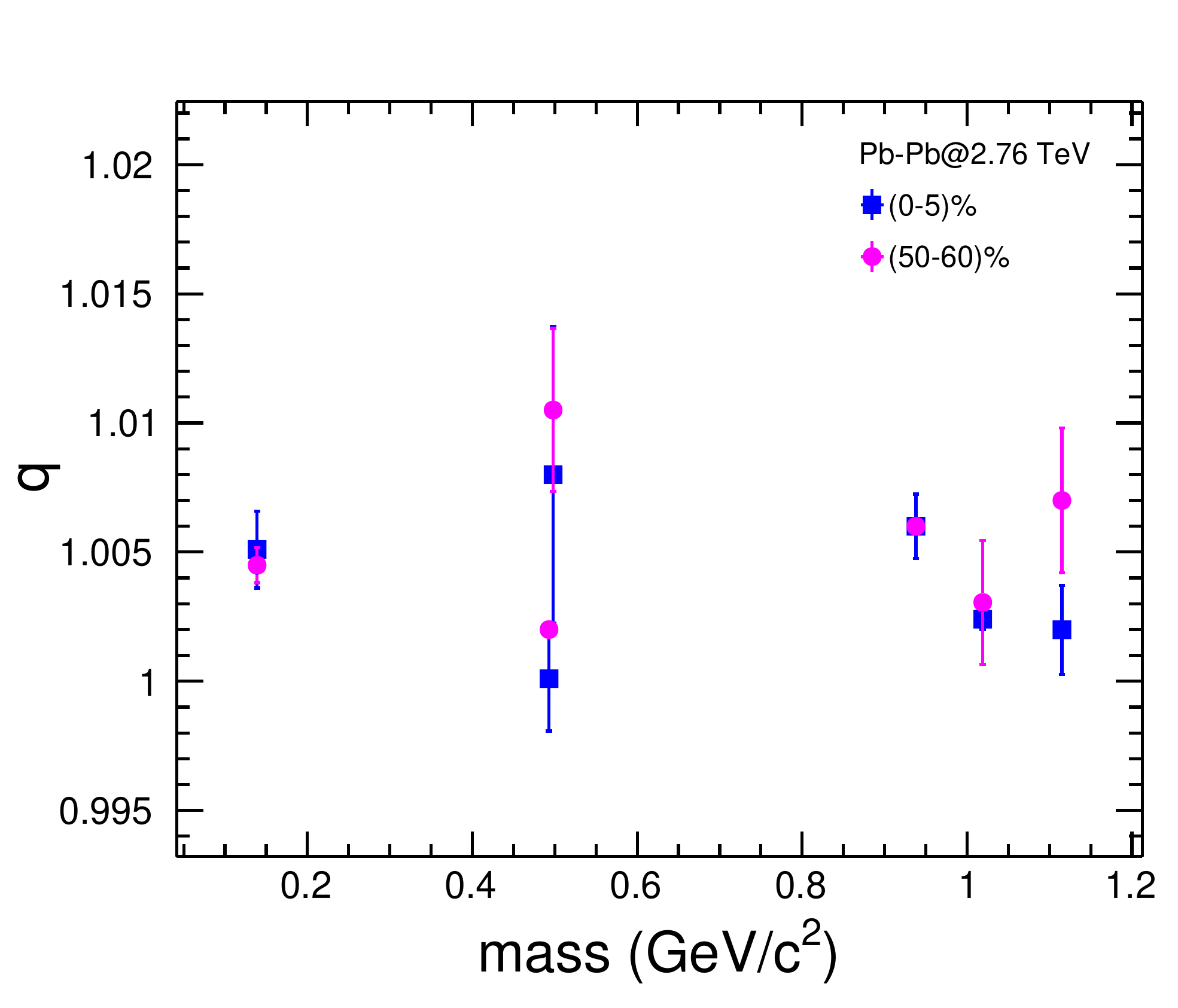}
\caption[]{(Color online) Non-extensive parameter, $q$ for identified hadrons.}
\label{fig12}
\end{figure}

In Fig.~\ref{fig12}, we have shown the mass dependence of non-extensive parameter, $q$. The parameter is an indirect measure of non-equilibration of any particle after evolving through thermal bath. Therefore, it is evident from the figure that different hadron species would show different $q$ values. From this figure we find that with increase in mass, the values of $q$ seem to decrease. Earlier in the discussion, while referring to Fig.~\ref{fig9}, we found that the value of $t_f/\tau$ increases with increase in mass. However, in a scenario, where $t_f$ is chosen the same for all particles through a choice of fixed kinetic freeze-out temperature, this translates to a lower value of relaxation time for high mass particles. In a kinetic theory, this is an indication that higher mass particles tend
to equilibrate early in time. This is truly reflected from Fig.~\ref{fig12}, where we do observe a decrease in the non-extensive parameter with increase of particle
mass. Furthermore, for a given mass, the $q$ values are higher for peripheral collisions in comparison with
central collisions. This is also an indication that peripheral collisions have higher tendency to go out of equilibrium. This behaviour of $q$ with centrality is in contrast with that of $t_f/\tau$ in Fig.~\ref{fig9}.

\section{Summary}
We have used Boltzmann transport equation in relaxation time approximation with non-extensive Tsallis statistics for the first time to estimate elliptic flow, $v_2$ for the identified hadrons in Pb+Pb collisions at $\sqrt{s_{NN}}$ = 2.76 TeV. The important findings of this work could be summarized as follows:
\begin{enumerate}
\item  We have observed that a formalism with BGBW alone, could not explain $v_2$ data beyond $p_T>$ 1.5 GeV/c. In order to improve upon this to describe $v_2$ for higher $p_{\rm T}$ values, we have incorporated non-extensive Tsallis statistics as the particle distribution in $p+p$ collisions as an input to BTE and evolve 
it to final distribution with BGBW as the equilibrium distribution. The present formalism could explain $v_2$ data for identified hadrons up to $p_{\rm T}$ = 5 GeV/c. 

\item The present formalism successfully connects the particle production in hadronic collisions with nuclear collisions at the LHC energy.

\item We have found a correlation between the radial and anisotropic part of the transverse flow while explaining the $v_2$ spectra in different centralities. In central collisions, the radial part is higher than the anisotropic part whereas the reverse is observed in the peripheral collisions. 

\item We have also observed that the ratio of freeze-out time to the relaxation time, $t_f/\tau$ has a contrasting behaviour with the non-extensive parameter, $q$ in different centralities.

\item In heavy-ion collisions, the peripheral collisions, compared to the central ones, have a higher tendency to go out of equilibrium. This is inferred from the observation of the non-extensive parameter, $q$ increasing towards peripheral collisions (deviating from the Boltzmann-Gibbs equilibrium value of $q=1$).

\item Hadrons with higher mass are found to have a greater tendency of equilibration. This is evident from the observation of monotonically decreasing relaxation time
with particle mass. This is also parallely supported by the observation of the non-extensive parameter, $q$ decreasing with the mass of the hadrons.

\end{enumerate}

\section*{Acknowledgements}
The authors acknowledge the financial supports  from  ALICE  Project  No. SR/MF/PS-01/2014-IITI(G)  of  
Department  of  Science  \&  Technology,  Government of India. ST acknowledges the financial support by 
DST-INSPIRE program of Government of India.

\end{document}